\documentclass[preprint,psfig]{aastex}

\usepackage{graphicx}

\newcommand{\be}{\begin{equation}}
\newcommand{\ee}{\end{equation}}
\newcommand{\nn}{\mbox{} \nonumber \\ \mbox{} }
\newcommand{\ba}{\begin{eqnarray}}
\newcommand{\ea}{\end{eqnarray}}

\newcommand{\Alfven}{ Alfv\'{e}n }
\newcommand{\curl}{{\rm curl\, }}
\newcommand{\E}{{\bf E}}
\newcommand{\B}{{\bf B}}

\renewcommand{\v}{{\bf v}}

\renewcommand{\div}{{\rm \,div\,}}

\newcommand{\Chandra}{{\it Chandra\,\,}}
\newcommand{\Bf}{magnetic field\,}
\newcommand{\Bfs}{magnetic fields\,}

\newcommand\etal{\textit{et al.\ }}
\newcommand\eg{\textit{e.g.\ }}
\newcommand\cf{\textit{cf.\ }}

\newcommand\lo{\mathrel{\raise.3ex\hbox{$<$}\mkern-14mu\lower0.6ex\hbox{$\sim$}}}
\newcommand\go{\mathrel{\raise.3ex\hbox{$>$}\mkern-14mu\lower0.6ex\hbox{$\sim$}}}
\begin{document}
\date{}  
\title{Magnetic draping of merging cores and radio bubbles in clusters of galaxies}
\author{M. Lyutikov
}
\affil{University of British Columbia, 6224 Agricultural Road,
Vancouver, BC, V6T 1Z1, Canada and
  Department of Physics and  Astronomy, University of Rochester,
     Bausch and  Lomb Hall,
         P.O. Box 270171,
	      600 Wilson Boulevard,
	            Rochester, NY 14627-0171, USA }

\begin{abstract}
Sharp
fronts observed by \Chandra satellite between  dense cool  cluster cores moving 
with  near-sonic velocity  through the hotter intergalactic  gas,
require strong suppression of thermal conductivity across the boundary.
This may be due to   magnetic fields 
 tangential to the contact surface  separating the two plasma components. 
 We point out that a super-Alfvenic
 motion of a plasma cloud (a core of a merging
galaxy) through a weakly magnetized intercluster medium leads to 
  "magnetic draping", formation of  a thin, 
  strongly magnetized   boundary layer with a 
 tangential magnetic field. For supersonic cloud motion, $M_s \geq 1$,
 \Bf inside the  layer 
  reaches near-equipartition values with thermal pressure.
 Typical thickness of the layer is 
$\sim L /M_A^2 \ll L$, 
where $L$ is the size of the obstacle (plasma cloud) moving
 with \Alfven Mach number $M_A \gg 1$. 
 To a various degree, magnetic  draping occurs both for sub- and supersonic
 flows, random and  ordered magnetic fields
 and it does not require plasma compressibility.
The strongly
 magnetized layer will 
 thermally isolate the two media and may contribute
 to the Kelvin-Helmholtz stability of the interface. 
Similar effects occur for radio bubbles, 
quasi-spherical expanding cavities blown up by AGN jets; in this case
the thickness of the external
magnetized layer is smaller, $\sim L /M_A^3 \ll L$.
\end{abstract}

\section{Introduction}
\Chandra observations of intergalactic medium (IGM) in clusters of galaxies 
 often show sharp discontinuities  in  gas density,
that separate dense cool gas moving with near-sonic velocity through the hotter
gas  \citep{Markevitch00,Vikhlinin}.
These fronts result from cluster mergers, when a cold subcluster core moves through
a hot IGM at transonic velocities. The front is very sharp.
For example, in the case of Abell A3667 the thickness of the front may be as thin as
$\sim 5 $ kpc, \citep{Vikhlinin}.

This   sharpness is surprising since heat conduction
 together with development
of the Kelvin-Helmholtz instability from
tangential motion of gas are expected to result in much broader transitions.
Subtle geometrical and kinetic effects may stabilize the contact \citep{ChurazovInogamov04} 
but this does not alleviate the problem of heat conduction.
For typical cluster parameters,
the Spitzer mean free path in unmagnetized plasma is
$\sim 30 $ kpc, much larger than  the thickness of the transition layer.
Thermal conductivity in  unmagnetized gas should lead to cloud 
evaporation on time scale of $\sim 10^7$ years \citep{EttoriFabian}.
Suppression of conductivity due to saturation of 
heat flux, when electron mean free path is comparable to the scale 
length  of the temperature gradient \citep{CowieMcKee}, is not sufficient. This 
 prompted  \cite{EttoriFabian} 
  to argue that
  classical conductivity has to be reduced  by a factor of between 250 and 2500.

Magnetic field  
 is the prime suspect in reducing the conductivity across the front, since Larmor radius, typically in 
 thousands of kilometers,
 is many orders of magnitude smaller than both electron mean free path 
 and the scale of the temperature gradient.
It was suggested that fields turbulent  on small scales may do the job, reducing Spitzer conductivity 
by a large factor \citep{Chandran}.
On the other hand, \cite{MedvedevNarayan} argued that
 that if  the scale of magnetic turbulence is smaller than the mean free path,
 and if the fluctuation spectrum extends over several decades in wave-vector,
 thermal conductivity is strongly enhanced  almost up to the Spitzer value.   
A possible caveat in this argument in application to cold fronts
is that it assumes that \Bfs in the
two media  do  intertwine.
 \cite{Vikhlinin}
 suggested that
 cross-boundary drift, heat conduction and KH  instabilities are all suppressed by  a large,
$ \geq 10 \mu G$,  \Bf along the contact surface. 
Such large, near-equipartition (with thermal pressure) 
\Bfs  must be local
to the front in view of the estimates of sub-microgauss fields by 
 Faraday rotation  measurements in 
radio sources
seen through the IGM \citep[\eg][]{KimKronberg91}, \citep[see reviews by][]{carillitaylor02,Ensslin05,Govoni06}.
 \cite{Vikhlinin} suggested that  a  \Bf parallel  to   the contact surface arises due to 
 shearing of an initially turbulent \Bf.

In this letter we discuss a straightforward dynamical effect
that leads  to formation of a narrow layer of  tangential, 
  near-equipartition 
\Bf at the  contact discontinuity.
Such region forms regardless of however small the \Bf is in the bulk of a cluster.
 This  effect is  well-known  in space  physics,  and 
sometimes is called magnetic draping or magnetic barrier. It is mostly pronounced for the interaction of  the Solar wind with
Venus and Mars, planets that do not have their own \Bf. It is also related  to the so-called 
plasma depletion layer, where
 the plasma density is depressed with
respect to values in the rest of the magnetosheath \citep{ZwanWolf76,Paschmann78}. 



\section{Properties of the transition layer}

\subsection{Formation: divergence of a \Bf on the contact in kinetic approximation} 

To explain the effect of magnetic draping in its simplest form,
consider an interaction of a  cloud moving through a 
{\it  weakly magnetized} medium parametrized by the ratio of  total
plasma to magnetic pressure
$\beta = 8 \pi p /B^2 \gg 1$. 
 Since the \Bf is weak, one may be prompted
 to neglect its dynamical effects on the flow completely.
 It turns out, that such approximation is not self-consistent and should break down close to the
 contact discontinuity separating  two fluids.

\begin{figure}
\includegraphics[width=0.95\linewidth]{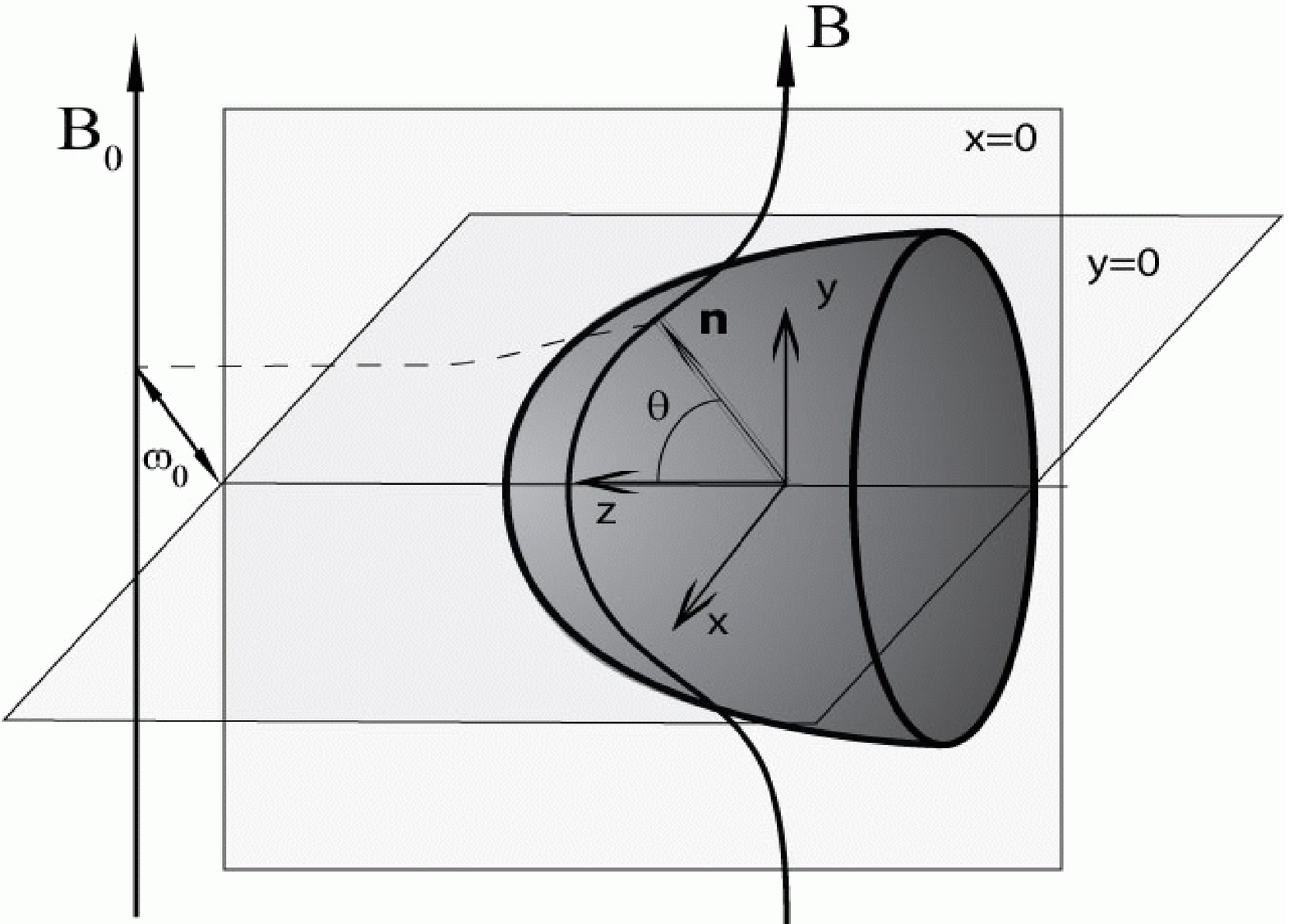}
\caption{Draping of \Bf lines around an axially symmetric obstacle. 
${\bf B}_0$ is magnetic field at infinity along $y$ direction, a cloud is moving along $z$ direction, dashed line is a flow line of a fluid element, 
$\varpi_0$ is the
initial distance from the symmetry axis, $\theta$ is a polar angle between a unit radius-vector from the origin  ${\bf n}$ to the current position
of a fluid element and $z$ axis.
}
\label{cluster}
\end{figure}

As a first step, which will be shown to be faulty, let us consider \Bf in kinetic
approximation,
when dynamical effects 
of the \Bf on  the flow are neglected  and the field lines are just advected with the flow satisfying
froze-in condition.
For simplicity, let us first assume that \Bf in bulk is large scale and is 
directed along $y$ axis, orthogonally to the
direction of cloud motion in $z$ direction (see Fig. \ref{cluster}).
In addition, we assume that the cloud is axially symmetric around the
stagnation line.
The motion of \Bf generates an inductive electric field, which under ideal approximation is
$\E = - \v \times \B $.
Combining induction equation
$\partial_t \B=-\curl \E$ with continuity equation
$\partial_t \rho + \div \rho \v=0$ gives
\citep{Alfven42}
\be
\left(\partial_t + (\v \cdot \nabla) \right) { \B \over \rho} =
{d  \over dt} { \B \over \rho}=
\left( { \B \over \rho} \cdot \nabla \right) \v
\label{vdb}
\ee
On the other hand if $\delta {\bf l}$ is an infinitesimal vector connecting two fluid
elements along the flow line,
its evolution with time is described by the same equation
${d  \over dt}  \delta {\bf l} = \left( \delta {\bf l} \cdot \nabla \right) \v$.
This shows that quantity ${ \B \over \rho}$ evolves according to 
the length along flow line.
Now, it is easy to see that as  field line is wrapped around the contact, its
length increases in proportion to the radius of the magnetosphere  $\varpi$
at a given location, 
${B / \rho}\sim ( {B / \rho})_0( \varpi/ \varpi_0)$ where
$  \varpi_0$ is the initial "impact parameter" of a given field line. 
On the contact $  \varpi_0 \rightarrow 0$ and as a consequence the ratio
$({B / \rho})  \rightarrow \infty$ \citep{PudovkinSemenov85}.

To understand this   in a  different way, note that  Eq. (\ref{vdb}) can be solved using Cauchy's
integral \cite[\eg][]{Stern66}:
\be
{\B \over \rho} = {\B_0 \cdot \nabla^{(0)} {\bf r} \over \rho_0}
\label{Cauchy1}
\ee
where $\nabla^{(0)} {\bf r}$ is a derivative of the coordinate of a flow 
 element ${\bf r}({\bf r}_0,t)$ with respect to initial coordinate
${\bf r}_0$, ${\bf B}_0$ is \Bf at infinity. In curvilinear coordinates with scale factors
$h_i$ Eq. (\ref{Cauchy1}) becomes
\be
{B_i \over \rho h_i} = {B_{0,j} \over \rho_0 h_j^{(0)}} { \partial x_i \over 
\partial x^{(0)}_j}
\label{h}
\ee
(summation on $j$ only).
For an axially symmetric flow  the scale factor $h_\phi^{(0)}$
corresponding to the cyclic variable $\phi$
is   proportional to $\varpi_0$,
initial radial cylindrical coordinate, which is  equal to the 
distance from the symmetry axis.
For toroidal component of \Bf,  Eq. (\ref{h}) gives
\be
{B_\phi   \over  \rho } = \left({B_{\phi} \over \rho} \right)_0  \left({\varpi \over
\varpi_0}\right)
\label{h2}
\ee
 where we used the fact that the azimuthal angle remains constant,
$\phi_0 = \phi$, so that  ${ \partial \phi \over \partial  \phi_0} =1$.
Eq. (\ref{h2}) shows that $B_\phi$
tends to infinity on the contact surface corresponding to $\varpi_0 \rightarrow 0$.

Divergence of the toroidal component of \Bf
 can also be shown explicitly by rewriting $\phi$ component of the 
Eq. (\ref{vdb}) (in spherical or cylindrical coordinate system) as
\be
\left( \partial _t + ({\bf v} \cdot \nabla) \right)
{B_\phi  \over \varpi \rho }=0
\ee
implying that  $ {B_\phi  / \varpi \rho }$
 remains constant along the flow lines and is equal to  $ {B_{0,\phi}
 / \varpi_0 \rho_0 }$ (\eg, for a constant \Bf at infinity along $y$ direction
 $B_{0,\phi} = B_0 \sin \phi$).

To illustrate the divergence of the other component of the \Bf parallel to the 
 contact (besides $B_\phi$)
 we will 
 assume a particular   form of the contact and a type of the fluid motion.
As an example, consider  a
 spherical body of radius $R_0$
 moving with velocity ${\bf v}_0$
 subsonically through an incompressible
fluid. The  velocity flow in the frame associated with the body  at a point defined by unit vector
${\bf n}$ is given by
${\bf v} = R_0^3( 3 {\bf n} ( {\bf n} {\bf v}_0 - {\bf v}_0))/(2 r^3) -{\bf v}_0$
where $r$ is a distance to the center of the  body
\citep[\eg][]{LLIV}. We can find an equation for flow lines
 $\varpi/\varpi_0 = 1/\sqrt{1-(R_0/r)^3}$, which gives
$\partial \theta/\partial \theta_0 = (\varpi / \varpi_0) (z_0/z) \sim \theta / \theta_0$
(last relation assumes small $\theta$ and $\theta_0$). Here $\theta$ is a
polar angle 
of a spherical system of coordinates aligned with the direction of the motion, Fig. \ref{cluster}.
 Thus, $\partial \theta/\partial \theta_0$ diverges
on the contact $\varpi_0 \rightarrow 0$ (or  $\theta_0 \rightarrow 0$) implying that 
the  component $B_\theta $ of the \Bf diverges on the contact.
Explicit calculations of the \Bf structure for this problem confirm this conclusion
\citep{BernikovSemenov}.
Incompressible motion past an axially symmetric body of arbitrary shape
can be obtained from the solution for a spherical body through conformal 
mapping of the boundaries.

It is easy to see that the divergence of $\partial \theta/\partial \theta_0$
at $\theta_0 \rightarrow 0$ occurs more generally than just in the case of
an incompressible flow considered above. It follows from the fact that
near  stagnation line  radial (in cylindrical coordinates) velocity
increases linearly with distance from the line, $v_\varpi \propto \varpi$, which implies
that $\partial \theta/\partial \theta_0 \sim \theta / \theta_0$.

The above derivations simplify along the stagnation line $\varpi =0$.
Then  the induction equation gives $ B_\varpi v_z=const=B_0 v_0$.
  Since at the stagnation point $v_z =0$, it follows that $ B_\varpi \rightarrow
  \infty$.

So far we have assumed that there are no discontinuities (shocks) in the flow.
In case of  
a supersonic motion the toroidal component of the \Bf
will experience a jump at the shock and, in addition, the flow lines will generally 
experience a bend. This will introduce correction factors to the above relations
but will not remove the divergence of a  \Bf on the contact.
The subsonic  incompressible flow considered above to  prove the divergence of the 
tangential to the contact component of a \Bf
 should be a reasonable
approximation near the  contact surface far downstream of a possible  forward shock.

The derivations above show that {\it \Bf is dynamically amplified  on the contact}. This is 
generally {\it independent} of whether the motion is supersonic or subsonic and is 
independent of plasma compressibility, though details of \Bf  amplification
will be somewhat different in these cases.
In addition, since governing equations can  be written in terms of total
derivatives along flow lines and thus are independent on the global structure of  a
\Bf, tangled fields  are  also subject to the same 
amplification of the component parallel to the contact. 
The amplification 
occurs due to longitudinal stretching of \Bf lines.
Thus, a  magnetized  boundary
 layer is created  in which \Bf may reach  near-equipartition  values
 (for supersonic bulk motion) regardless of 
 its value in the bulk (it should be non-zero, though).
This effect is called magnetic draping or magnetic barrier.

Note  also that the 2-D case is very different from 3-D case. 
For example, in 3-D, the induction equation for stationary flow 
gives 
\ba &&
({\bf v} \cdot \nabla) (r \sin \theta  E_\phi)=0
\ea
so that the quantity $ \varpi E_\phi
$ 
is constant along the flow line, $ \varpi E_\phi = \varpi_0 E_{0,\phi}$.
\footnote{Tangential component of  electric field is continuous across tangential discontinuities, like shocks.}
 This  implies that toroidal 
 electric field vanishes on the contact,  $\varpi_0 =0$.
Contrary to this, in 2-D
 the electric field is constant everywhere, including on the contact,
 $\E = (B_l v_n-B_n v_l) {\bf e}_x$, where coordinate $ l$ is
 along the generators of the contact surface from the axis of the symmetry
 and $n$  is orthogonal to it.
 Divergence of \Bf on the contact then follows from the
 requirement that normal components of both the velocity and \Bf should vanish,
 $v_n = B_n =0$,  implying  $ B_l \rightarrow  \infty$
 \cite[\cf][model for planar flows]{Parker73}.

For super-Alfvenic motion, $M_A \geq 1$,
the only case when a magnetized layer does not form near the contact is when
a cloud moves along the external \Bf. In this case both $B_\phi =0$ and
$E_\phi =0$ everywhere.
In an axially symmetric, incompressible, stationary  flow,
velocity and \Bf may be expressed in  terms of 
flux functions $\Psi$ and $P$,
 ${\bf v} = \nabla \Psi / \varpi$, $\B= \nabla P/ \varpi$ where
 $P= A_\phi \varpi$ and  $A_\phi$ is a toroidal component
 of vector potential.
Condition $E_\phi =0$ then gives (\eg in spherical coordinates)
\be
\partial_\theta P \partial_r \Psi - \partial_\theta  \Psi  \partial_r P =0
\ee 
implying that $ P= f(\Psi)$ and 
$B_r = v_r f'$, $B_\theta= v_\theta f'$.
In particular, if \Bf at infinity
is constant in space then $ f'= $constant, so that two flux functions are linearly
related, $  P \propto \Psi$ . Since for a spherical obstacle
$v_r=0$ and $v_\theta$ is finite on the contact, \Bf
remains finite everywhere.

 The divergence of \Bf on the contact is, of course, a consequence of the kinetic approximation.
In full MHD case \Bf is expected to be amplified to  the  point when
its dynamical influence cannot be neglected anymore. For sonic and supersonic
motion this corresponds to 
near-equipartition (with total pressure, 
not just those of relativistic particles) fields.
The layer itself must be modeled using MHD, 
not fluid equations. Inside the layer 
non-isotropic magnetic pressure  will break the axial symmetry of the flow
even for an axially symmetric obstacle.
Stagnation point flow changes into stagnation line flow, 
where stream lines branch off
not only at the stagnation point,
but also at all points along the symmetry line ($x=0$ plane in Fig. \ref{cluster}).
In case of a flow of plasma cloud,
since position of the contact is determined by the
pressure balance between two media, this will lead to deformation of the contact surface
(by a fraction $\sim 1/M_A^2$).

A magnetized layer should also be associated with depletion of plasma.
To see the reason for this, note that
the total pressure, which is a sum of magnetic and gas pressures, should 
remain approximately constant well behind a possible forward shock
(neglecting possible temperature anisotropy, see below). Increasing
 magnetic  pressure requires a decrease in  gas pressures, which in a  polytropic  
 fluid with adiabatic index $\gamma$ (\eg for  isentropic ideal gas)
 is accompanied by a decrease in density (approximately as
 $\rho  \propto (\beta/(1+\beta))^{1/\gamma}$, decreasing with decreasing 
 $\beta$).
Similarly, the plasma temperature should decrease as well,
$T  \propto (\beta/(1+\beta))^{(\gamma-1)/\gamma}$.
 Thus, for a supersonic motion, 
the plasma density first
increases at the forward shock  and then decreases 
 inside the magnetic barrier close to the contact
\citep{wu92}.

Reality is a bit more complicated than this simple estimate since several competing
 plasma physics effects come into play. First,
compression of plasma in a \Bf creates  anisotropy of plasma pressure
due to conservation of adiabatic invariants
\cite[double adiabatic model of][]{Chew}. 
When compressed normally to \Bf lines, 
conservations of the  adiabatic invariant $p_\perp/(B \rho)=const$ and
effective   polytropic   index $\gamma_\perp =2$
would lead to
density and transverse  temperature  $T_\perp$ increasing with a magnetic field.
But strong plasma
anisotropy leads to expulsion of plasma from high \Bf regions due to mirror forces
(magnetic bottling, reflection of particles from regions of high \Bf, is the best known example
of mirror forces). Second,
onset of plasma instabilities may limit the pressure anisotropy 
(\eg to $T_\perp \sim T_\parallel$). In case of planets interacting with the Solar wind,
typically isotropic MHD model give a reasonable fit to \Bf, pressure and
temperature
profiles \citep{DentonLyon96,Pudovkin99,SongRussell02}.
Behavior of the magnetized boundary layer  in some aspects is opposite to the purely  fluid
case. For example, the former leads to density minimum on the contact while 
the later predicts density maximum \citep{Lees64}, \citep[see][for a recent review]{SongRussell02}.

Historically,
magnetic draping effect 
was  somewhat  a  surprise in modeling of Solar wind interaction with  planets. 
It was expected that for small magnetization the flow may be computed from purely 
hydrodynamical equations, and a \Bf may be added later using frozen-in condition
 \citep{Spreiter66}.
Using this prescription
\cite{Alksne67} found that the \Bf goes to infinity at the contact, especially  strongly
at plane containing the symmetry axis and \Bf ($y=0$ plane in our notations).
\cite{ZwanWolf76} \citep[see also][]{SouthwoodKivelson95}
calculated in details the properties of the magnetic barrier
and predicted that it should be associated with depletion of plasma density, as 
\Bf lines are stretched along the contact surface and plasma is allowed to
flow along stretched \Bf lines and, in
addition, developing of temperature anisotropy leads to magnetic mirror forces pushing plasma away
from regions of high \Bf.

These theoretical ideas have been generally confirmed by direct satellite observations
of the Terrestrial \citep[\eg][]{Paschmann78,Crooker79,kallio94,wang03}, 
Cytherean  \citep{Biernat99} and 
Martian \citep{oieroset04} depletion layers.
Overall, observations seem to be  consistent with modern full MHD models
\citep[\eg][and preceding references]{kallio98,erkaev00}. Magnetic draping also occurs  at the outer heliosphere and may be related to low frequency, $\sim 3 $ kHz, radiation observed by the Voyager
spacecraft \citep{Cairns04}.
In astrophysical setting these ideas have been touched upon theoretically by
\cite{Kulsrud,Rosenau76} for the case of supernova expansion and by \cite{lyutikov}
 in the case of  relativistic GRB outflows.
A number of numerical experiments also saw formation of the magnetic barrier
\citep{MacLow94,Jones96,Gregori00,Asai04,Asai05},
producing  results in agreement with the above theoretical estimates.
In particular, in case of merging cluster cores,
the work of \cite{Asai05} clearly shows formation of magnetic barrier and its larger thickness
in 2-D, as expected (see section \ref{thickness}).

\subsection{Thickness of the magnetic barrier}
\label{thickness}

For low magnetization, $\beta \gg 1 $, a flow may be separated in
two regions: a  bulk, where motion is nearly hydrodynamic,
and a boundary layer, where effects of a \Bf are important. Flows in the two regions
must be matched at their boundaries. Thus, magnetic draping almost does not 
affect, for example, the location of the forward shock in front of the obstacle.

Next we estimate a
thickness of the transition layer.
Both for subsonic and supersonic motion of the cloud
through IGM, the motion near the critical point is strongly subsonic and can be considered incompressible.
In this case the velocity field is $v_\varpi = - ({3 V_0/2}) ({\varpi/L})$, $v_z = {3 V_0 } ({z/L})$
\citep{LLIV}.
Then, along the stagnation  line $\varpi=0$ \Bf evolves according to
\be
B_\varpi+ 2 z \partial_z  B_\varpi =0
\label{Bvar}
\ee
giving 
$B_\varpi \propto 1/\sqrt{z}$. 
To estimate 
  the  thickness of the magnetized layer,
 note that inside the layer 
magnetic pressure becomes of the order of
ram pressure, $B^2 \sim 8 \pi \rho v_0^2$ (at this point magnetic forces
will strongly affect the plasma flow).
 If the typical size of the
plasma cloud  is $L$ and 
$M_A = v_\infty / v_{A,\inf}$ is \Alfven  Mach number defined in terms of \Alfven velocity $v_{A,\inf}$
at infinity,
then, using Eq (\ref{Bvar}) we find
\be
{\Delta r \over L} \sim {1\over M_A^2} \ll 1
\ee
The value of plasma $\beta$ inside the magnetized sublayer is 
$\beta_{in} \sim (1 +M_s^2))/M_s^2$, where 
 $M_s= v_\infty /c_s$ is sonic  Mach number at infinity.
Thus, for supersonic motion,
$M_s \geq 1$, a near-equipartition layer, $\beta \sim 1$, forms
\citep[see also][]{ZwanWolf76}.
In case of subsonic motion plasma $\beta$ inside the 
 sublayer is  much smaller than in the bulk, 
 $\beta_{in}/\beta \sim 1/M_A^2 \ll 1$ (so that plasma is more strongly magnetized). 
Both for subsonic and  supersonic motion,
the thickness of the magnetized boundary 
layer is much smaller that the size of the cloud  (for $M_A \geq 1$). 
Note, the dimensionality of the problem is an important issue.
Repeating the above estimates  for a 2-D flow, the thickness of the magnetized layer  is 
$ \Delta r/L \sim 1/M_A$ \citep{Erkaev95}, much larger than in the 3-D case.

The time it takes to form the layer may be estimated from a condition that
a swept-up magnetic flux is of the order of the magnetic flux through the layer.
A layer forms rather  quickly, after the cloud traversed a length
$l \sim L / M_A \leq   L$. An exception is when the cloud moves very slowly, 
sub-Alfvenically $M_A \leq 1$, in which case  no magnetized layer forms
anyway.



\section{Application to cluster cold fronts}

A well-studied case is Abell 3667, which we use below for numerical estimates
\citep{VikhlininMarkevitch02}. In this case the 
density of the hot component is    $n_h  \sim 8  \times 10^{-4}$ cm $^{-3}$, its temperature $T_h \sim 8 $ keV,
 front velocity  $\sim 1500 $ km s$^{-1}$,  corresponding to  sonic Mach number $
 M_s \sim 1$.
 The Spitzer Coulomb free path is  $\lambda_e \sim 30$ kpc \citep{EttoriFabian}.
If an average \Bf in the cluster 
is $\sim 1 \mu$G, the 
electron Larmor radius is only
$\sim  4  \times 10^8$ cm. 
Thus, the Larmor radius is many orders of magnitude smaller than the collision length, so that
 even in case
 of highest possible cross-field diffusion (Bohm diffusion), cross-field conductivity is virtually zero.

Assuming 
 $B =1 \mu$G, 
the  plasma $\beta$ parameter in the bulk is  then
$ \beta \sim 400 $, 
\Alfven Mach number is $M_A \sim \sqrt{ \gamma \beta/2} M_s \sim 20$,    so that 
for the core of $L \sim  500 kpc$ the 
thickness of the strongly magnetized  boundary layer is  $  \sim L/M_A^2 \sim 1.25 kpc $. 
This is somewhat  smaller than 
 \Chandra resolution for a typical cluster, but since  the thickness
depends strongly on the assumed IGM \Bf, $\propto B^2$, 
it is possible that in some cases the layer may be resolved.

If the IGM is permeated by large scale \Bf, 
the magnetized layer provides a contribution to rotation
measure of the order of
\be
RM  \sim 15 \, {\rm rad/m}^2 \, \left({M_s \over 1}\right)^{-1}\,
\left({ B_\infty \over 1 \mu {\rm G}} \right)^{2} 
\left( {n \over 8 \times 10^{-4}} \right) ^{1/2} 
\left( {T \over 8 {\rm keV}} \right)^{-1/2}
\left( { L \over 500  {\rm kpc} }  \right)
\ee
This is a relatively small value. In addition,
since only the component of  \Bf  along the line of sight  contributes
to the  rotation
measure, this  estimate is 
 subject to
strong geometrical variations over the contact.

\section{Radio bubbles}
\label{Radiobubbles}

Magnetic draping should also be important for stabilization of  rising radio bubbles,
X-ray emission voids of up to 30 kpc in size,
against Rayleigh-Taylor and Kelvin-Helmholtz instabilities and, 
similar to merging cores, suppression of thermal conductivity across the
front. For example,  \citep{Robinson04} (see also \citep{Churasov01,Bruggen02,JonesDeyoung})
showed that in the absence of \Bf, bubbles are disrupted by 
hydrodynamic instabilities and effects of
thermal conduction. In order to stabilize the contact, {\it locally} \Bf should be of the  
order of equipartition field, while  the field in bulk is much smaller. 
The effect of magnetic draping provides exactly what is needed for stability and suppression of conductivity:
near-equipartition  \Bf tangential to the contact (in case of nearly sonic
or supersonic expansion).
Hydrodynamically, the problem of expanding 
cavity blown by an AGN jet is similar to the one considered by \cite{Kulsrud}
of a supernova remnant expanding into ISM \citep[see also][]{DeYoung02}. 

\begin{figure}
\includegraphics[width=0.95\linewidth]{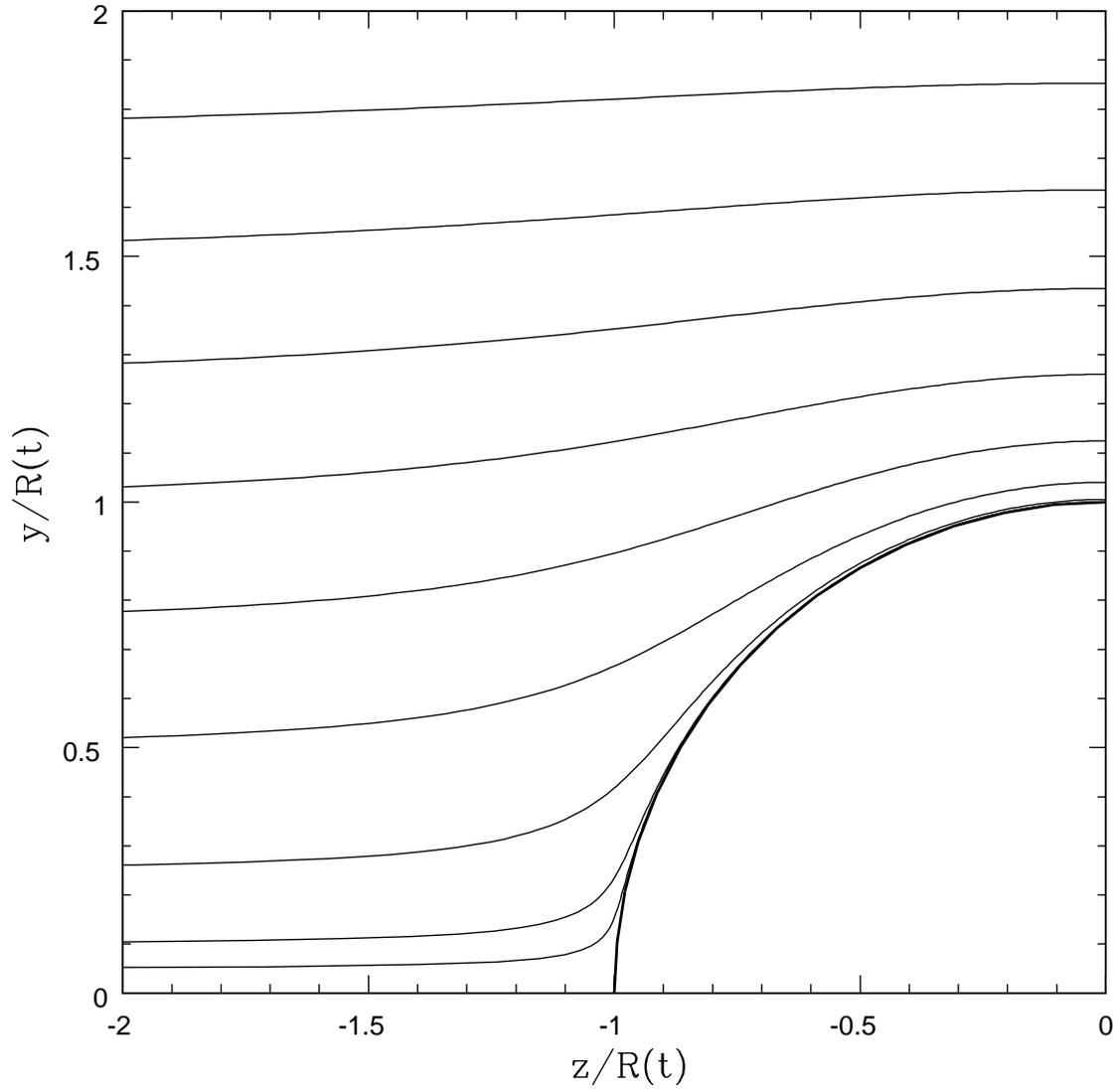}
\caption{Magnetic field lines in the $z-y$ plane for subsonic incompressible 
expansion of a sphere into a 
medium with a constant \Bf along $z$ axis at infinity, kinematic approximation.
Note the strong compression of field lines on the contact $r=R(t)$.}
\label{halo}
\end{figure}

Consider an impermeable  sphere  expanding radially  with radius $R(t)$ 
into incompressible medium permeated by constant \Bf $B_0$.
 From the incompressibility condition we find 
$v_r = R'(t) (R(t)/r)^2$. Inductive electric field is in $\phi$ direction and
the  induction equation gives
\ba &&
\partial_t B_\theta= {\partial_r ( B_\theta v_r r )  \over r}
\nn &&
\partial_t B_r = -  {\partial_\theta (\sin \theta  B_\theta  v_r) \over r \sin \theta}
\ea
Assuming that $B_\theta$ depends on self-similar variable $\xi =r/R(t)> 1$, 
 equation for $B_\theta$ becomes
\be
B_\theta' (\xi^3-1)+ {B_\theta \over \xi} =0
\ee
which gives
\ba &&
B_\theta = {  \sin \theta \over (1- \xi^{-3})^{1/3}} B_0
\nn &&
B_r = - \cos \theta (1- \xi^{-3})^{1/3} B_0
\label{BB}
\ea
 This implies that tangential component of \Bf diverges on the contact $\xi=1$ as 
$\propto (3 (\xi-1))^{-1/3}$. 
Using Eq. (\ref{BB}) we can find equation for field lines,
$(\xi^3 -1) \sin \theta=\varpi_0/R(t)$, see Fig. \ref{halo}.
Similarly to the case of translational motion, amplification of \Bf on the 
contact will lead to formation of a magnetized boundary layer which thickness now is
smaller by a factor $M_A$:
 $\Delta r / R \sim 
M_A^{-3}$
 \cite[\cf][]{Kulsrud}.
MHD simulations generally confirm this picture: \cite{Robinson04} and 
\cite{JonesDeyoung} find that
 modest IGM magnetic fields can  suppress thermal conductivity and
 stabilize the rising bubbles against disruption by 
 fluid instabilities. 

Presence of a large scale \Bf inside a bubble, a leftover from AGN pumping, may ease constraints 
on stability. An internal \Bf  may also be needed to compensate for high external pressure
since  bubbles  appear to be in pressure equilibrium, but the
absence of X-ray emission
argues for a lower temperature in comparison with external IGM. 
An additional stabilizing 
effect may come from a pile-up of \Bf {\it inside} the contact discontinuity. 
Typically, the entropy of  gas in the cores
is lower than at the outskirts, resulting in  motion of gas between the core and
the contact. This will similarly create a magnetized layer  
on the inside of  the contact. 

\section{Conclusion}

We point out  that presence of a  dynamically unimportant \Bf in the bulk of the IGM
leads to  formation of a strongly magnetized boundary layer which
undoubtedly affects the mechanical and thermodynamical coupling
between plasmas of the cold merging cores or rising AGN blown bubbles on the one side
and hot IGM plasma on the other side.
To a different degree this occurs both around sub-sonically and super-sonically
moving flows, for random and large-scale \Bfs and does not  require  plasma  
compressibility.
Primarily, magnetic draping leads to strong suppression of thermal conductivity
across the contact and may explain observed narrowness of the transition layers. 
A boundary layer of near-equipartition \Bf (for supersonic motion $M_s \gg 1$)
is also expected to  stabilize the  KH instability of the contact,
especially
close to the critical point of the flow (in case of merging cores).

Thus, even weak bulk  magnetization strongly affects the interaction of  a flow with an obstacle.
This runs contrary to expectations that a  weak \Bf  in the bulk  should not affect much
overall dynamics of a  merging cluster. 
 For example, 
Heinz et al. (2003) \citep[see also][]{NagaiKravtsov03,Bialek02} 
carried out hydrodynamical simulations of the interaction of cold
subcluster plasma and hot ambient matter neglecting heat conduction
and magnetic fields. 
When \cite{Asai04} repeated simulations of Heinz et al. (2003),
they found  that
when the Spitzer
conductivity is adopted, 
the subcluster evaporates rapidly and the cold front is not formed.



The main prediction of the model is 
that \Bf may reach near-equipartition values
and be directed along the 
contact separating  two fluids. This is best tested with 
high resolution polarization radio observation.
If high-energy non-thermal
particles are accelerated locally and produce synchrotron emission,  
the direction of the magnetic field may be then
inferred from linear polarization. 
For example, radio observations of  NGC 4522, a prominent galaxy in the Virgo cluster, 
indeed show \Bfs along the front \citep{Vollmer04}.
Parallel magnetic fields  are also observed
in  the case of peripheral
 features seen in  Abell 2256 \citep{Rottgering94,Clarke02}.
In addition, a magnetized layer may contribute to the rotation measure of background radio sources.
Naturally, in both cases we can estimate only average values of \Bf
along the line of sight and it's not
easy to single out contributions from a narrow, strongly magnetized layer.
 

A number of observations may already be interpreted in 
the framework of the model: enhanced RMs are often  seen at the edges of 
structural features in clusters.
E.g.,  \cite{Carilli88} see enhanced RMs at the edges of hot spots 
in Cygnus A  \citep{Carilli88}, 
\cite{taylor92} find an enhanced RM
at the edge of one of the
hot spots in 3C 194.
 Though our estimates
 in application to Abell 3667 show that variations of RM across the front are
 typically small (and even smaller for expanding bubble
 due to smaller thickness in that case, Section \ref{Radiobubbles}), total RM is a strong function of assumed \Bf in the bulk, $\sim B^2$.
The Expanded Very Large Array (EVLA) will be most instrumental in mapping 
\Bfs in clusters of galaxies.

Another prediction of the model is 
that due to the depletion of plasma from the magnetized sheath
we may observe a 
 narrow layer $\sim 
1 kpc$ thick of {\it suppressed X-ray emission} on the outside edge of cold
fronts. Naturally, observations needed to test this are very challenging.

Dynamic amplification of an external \Bf on both sides of the contact creates conditions favorable for
reconnection between external and internal \Bfs 
\cite[\eg][]{pudovki2002}. As a result of reconnection, 
normal components of velocity at the contact surface
may be non-zero, determined either by stripping or by  physics of a 
reconnection layer.
Reconnection is known to be an efficient site of particle acceleration which 
 may produce observable radio and non-thermal
X-ray signals \citep[\cf a  model 
of non-thermal emission in young supernova remnants of ][]{LyutikovPohl}.
Wrapping of \Bf lines may create conditions favorable for reconnection in the
 wake of the core, similar to reconnection in the Earth magneto-tail \citep{Galeev}.





Numerical simulations required to correctly
describe the structure of the magnetic barrier are bound to be complicated.
First, they must be done in full 3-D using MHD approximation: fluid models lead to qualitatively
different results, while 2-D MHD simulations grossly overestimate the thickness of the magnetic barrier
(by  a factor $M_A \gg 1$ if compared with the 3-D case). In addition, 
since it is expected that plasma inside the  magnetic barrier becomes anisotropic,
simulations should account for possible temperature anisotropy, \eg 
within the framework of Chew-Goldberger-Low  \citep{Chew} theory. 
Finally, kinetic effects, like self-limiting
temperature anisotropy, may be important as well.

Finally we note that the effect of magnetic draping may be important in other astrophysical
applications, 
in particular in case of suppressed conductivity between dense cold HI ISM clouds
and a lower density warmer medium \citep[\eg review by ][]{Bregman04}.
Supernova shocks and   outflows from OB associations  induce 
a large scale motion 
in  the warm medium which will lead to magnetic blanketing of cold clouds.
This effect should be especially important for  High-Velocity Clouds.


I would like to thank Andrei Kravtsov, Jean Eilek, 
Maxim Markevitch, Christoph Pfrommer, Dmitry Uzdensky and Alexey Vikhlinin for comments and discussions.

\end {document}